\documentclass[12pt,aps,floatfix]{revtex4}

\usepackage{graphicx}
\usepackage{amssymb}
\usepackage{amsmath}
\usepackage{amsfonts}
\usepackage{epstopdf}
\usepackage{epsfig}
\usepackage{wrapfig}

\newtheorem{theorem}{Theorem}

\newtheorem{corollary}[theorem]{Corollary}

\newtheorem{definition}[theorem]{Definition}

\newenvironment{proof}[1][Proof]{\noindent\textbf{#1.} }{\ \rule{0.5em}{0.5em}}

\newtheorem{myproposition}{Proposition}

\newtheorem{mydefinition}{Definition}
\newtheorem{mytheorem}{Theorem}

\begin{document}

\title{A theorem on the quantum evaluation of Weight Enumerators for a certain class of Cyclic Codes with a note on Cyclotomic cosets}
\author{Joseph Geraci}
\affiliation{Department of Mathematics, University of Toronto, Toronto, ON M5S 2E4, Canada}
\affiliation{Chemistry Department, University of Southern California, CA 90089, USA}
\author{Frank Van Bussel}
\affiliation{Max Planck Institute for Dynamics and Self-Organization, Bunsenstr. 10, 37073 G\"ottingen, Germany}

\begin{abstract}
This note is a stripped down version of a published paper on the Potts partition function, where we concentrate solely on the linear coding aspect of our approach. It is meant as a resource for people interested in coding theory but who do not know much of the mathematics involved and how quantum computation may provide a speed up in the computation of a very important quantity in coding theory. We provide a theorem on the quantum computation of the Weight Enumerator polynomial for a restricted family of cyclic codes. The complexity of obtaining an exact evaluation is $O(k^{2s}(\log q)^{2})$, where $s$ is a parameter which determines the class of cyclic codes in question, $q$ is the characteristic of the finite field over which the code is defined, and $k$ is the dimension of the code.  We also provide an overview of cyclotomic cosets and discuss applications including how they can be used to speed up the computation of the weight enumerator polynomial (which is related to the Potts partition function). We also give an algorithm which returns the coset leaders and the size of each coset from the list $\{0,1,2,\dots,N-1\}$, whose time complexity is soft-$O(N)$. This algorithm uses standard techniques but we include it as a resource for students. Note that cyclotomic cosets do not improve the asymptotic complexity of the computation of weight enumerators.   
\end{abstract}

\maketitle

\section{Introduction}

There have been many quantum algorithms applied to problems of interest to mathematical scientists, including the famous Shor's algorithm for prime factorization \cite{Shor}, approximations of the Jones polynomial \cite{Wocjan:06,Aharonov:06}, approximations of the Tutte polynomial \cite{Aharonov:07} and algorithms for Gauss sums and Zeta functions \cite{van,Kedlaya:05}. In this paper we give an algorithm for the evaluation of another variant of the Tutte polynomial, the weight enumerator polynomial for classical codes, which is of interest to people in the engineering sciences and other branches of applied mathematics. In particular, we use an approach recently outlined in \cite{geraci} to allow quantum computers to return the exact weight enumerator polynomial for a certain restricted subset of linear codes. It relies on a quantum oracle that is able to run an algorithm for discrete log (as in \cite{Shor}) and an algorithm for the estimation of Gauss sums (or Zeta functions) which is known to be as computationally hard as the evaluation of discrete log \cite{van}. 
 
Cyclotomic cosets are a partition of the list $\{0,1,2,\dots,N-1\}$ into $m$ unique subsets. While doing research at the intersection of classical coding theory and statistical physics, one of the authors found that if one knew one element from each coset and the size of each coset, i.e., the number of members in that coset, then certain instances of a hard problem became computationally easier. Specifically, one is able to obtain the exact partition function for the Potts model over a specific family of graphs if one had access to a quantum oracle\cite{geraci}. This was due to certain algebraic symmetries that a  function that we needed to compute had, i.e., Gauss sums, which we review. This author also found that there was a lack of literature on computing cyclotomic cosets. (Note that \cite{mathworks} does provide a computational tool for finding cyclotomic cosets.) Here we provide a classical algorithm for the computation of an element from each set, the so called coset leader or coset representative, and also the number of elements in each coset.  The complexity of the algorithm is $O(N)$.

We also present a theorem on the exact evaluation of the weight enumerator polynomial for a certain family of codes using quantum computation. This theorem is actually a stripped down version of the main theorem presented in \cite{geraci}, i.e., with no mention of statistical physics. 

\section{Cyclotomic Cosets}

Let $S = \{0,1,2,\dots,N-1\}$ and let $p$ be prime such that $\gcd(N,p)=1$. The
p-cyclotomic cosets of this set is given by the collection of subsets 
\begin{equation*}
\{0\},\{1,p,p^{2},\dots ,p^{r}\},\dots ,\{a,ap,ap^{2},\dots ,ap^{s}\}
\end{equation*}%
where elements are computed mod N and $s$ is the minimal exponent such that $%
a(p^{s}-1)=0 \mod N$ i.e. $s$ is the smallest integer before one
begins to get repeats in the coset. (The same is true for $r$.)  

As an example consider $N=16$ and $p=3$.
One obtains 
\begin{equation*}
\{0\},\{1,3,9,11\},\{2,6\},\{4,12\},\{5,15,13,7\},\{8\},\{10,14\}.
\end{equation*}%

One sees that this defines an equivalence relation, i.e., for $g,f \in S$ we have that $g\sim f$ if $g=f \cdot p^l \mod N$ for some $l$. Each equivalence class in known as a cyclotomic coset or class and referred to as $C_j$ where $j$ is the coset leader, i.e., the smallest coset representative. For the example given above we have $C_0 = \{0\}$ (as always), $C_1= \{1,3,9,11\}$, $C_2 = \{2,6\}$, $C_4 = \{4,12\}$, $C_5 = \{5,15,13,7\}$, $C_8 = \{8\}$, and $C_{10} = \{10,14\}$.

\section{Applications}

\subsection{Factorization of $X^N-1$}

We present a few results with no proof. Proofs can be found in \cite{vermani}. Take $p$ to be prime.

\begin{definition}
Consider an element $\alpha$ in the finite field extension $\mathrm{GF}(p^l)$ of $\mathrm{GF}(p)$. The minimal polynomial of $\alpha$ is the monic, irreducible polynomial $M(x)$ of least degree such that $M(\alpha)=0$. 
\end{definition}

The following is a classical result and it is an extension of the fact that $X^{p^n} - X$ is equal to the product of all monic polynomials, irreducible over $\mathrm{GF}(p)$ whose degree divides $n$. The idea is that once one has the cyclotomic cosets of $S$, then one can find a factorization of $X^N-1$ into a product of monic polynomials as well. 

\begin{theorem}
\[ M_s(X) \equiv \prod_{\eta \in C_s} X - \eta \] is the minimal polynomial of $\alpha^s$ over $\mathrm{GF}(p^k)$.
\end{theorem}

\begin{corollary}
\[ X^N - 1 =  \prod_s M_s(X) \] where $s$ runs over any set of coset representatives modulo N over $\mathrm{GF}(p)$. 
\end{corollary}
 
The above theorems provide a basis for the factorization of $X^N -1$ which has applications in the theory of error correcting codes. We provide some details in the next section. Detailed examples of using cyclotomic cosets for finding factorizations are provided in \cite{vermani}. 

\subsection{Cyclic Codes}

We go into some detail here for it will be useful background for a recently discovered theorem that we include in this paper. Let us recall some definitions from algebra after we define linear codes. Take $q$ to be prime or a power of a prime and let $\mathbf{F}_{q} = \mathrm{GF}(q)$

\begin{definition}
 A linear code $C$ is
a $k$ dimensional subspace of the vector space $\mathbf{F}_{q}^{n}$ and is
referred to as an $[n,k]$ code. The code is said to be of length $n$ and of
dimension $k$.
\end{definition}

\begin{definition}
A linear code $C$ is a cyclic code if for any word $(c_{0},c_{1},\dots
,c_{n-1})\in C$, also $(c_{n-1},c_{0},c_{1},\dots ,c_{n-2})\in C$. If $C$
contains no subspace (other than $0$) which is closed under cyclic shifts
then it is irreducible cyclic.
\end{definition}

\begin{mydefinition}
A ring is a set $R$ which is an abelian group $(R,+)$ with $0$ as the
identity, together with $(R,\times )$, which has an identity element with
respect to $\times $ where $\times $ is associative.
\end{mydefinition}

\begin{mydefinition}
An ideal $I$ is a subset of a ring $R$ which is itself an additive subgroup
of $(R,+)$ and has the property that when $x\in R$ and $a\in I$ then $xa$
and $ax$ are also in $I$.
\end{mydefinition}

\begin{mydefinition}
A principal ideal is an ideal where every element is of the form $ar$ where $%
r\in R$.
\end{mydefinition}

Thus, a principal ideal is generated by the one element $a$ and a principal
ideal ring is a ring in which every ideal is principal.

There is an important isomorphism between powers of finite fields $\mathbf{F}_{q}^{n}$ and a certain ring of polynomials. Let $
(x^{n}-1)$ be the principal ideal in the polynomial ring $\mathbf{F}_{q}[x]$ generated by $x^n-1$. 

Therefore the residue class ring $\mathbf{F}_{q}[x]/(x^{n}-1)$ is isomorphic
to $\mathbf{F}_{q}^{n}$ since it consists of the polynomials 
\begin{equation*}
\{a_0 + a_1x + \cdots + a_{n-1} x^{n-1} | a_i \in \mathbf{F}_q, 0 \le i <n
\}.
\end{equation*}

Taking multiplication modulo $x^n -1$ we can make the following
identification: 
\begin{equation}
(a_0,a_1,\dots ,a_{n-1}) \in \mathbf{F}_q^n \longleftrightarrow a_0 + a_1x +
\cdots + a_{n-1} x^{n-1} \in \mathbf{F}_q[x]/(x^n-1) .  \label{corr}
\end{equation}

This implies the following theorem.

\begin{mytheorem}
A linear code $C$ in $\mathbf{F}_{q}^{n}$ is cylic $\iff $ $C$ is an ideal
in $\mathbf{F}_{q}[x]/(x^{n}-1)$.\cite{Lint:book}
\end{mytheorem}

Note that $\mathbf{F}_{q}[x]/(x^{n}-1)$ is a principal ideal ring and
therefore the elements of every cyclic code $C$ are just multiples of $g(x)$, the monic polynomial of lowest degree in $C$; $g(x)$ is called the
generator polynomial of $C$.  We see that $g(x)$
divides $x^{n}-1$ since otherwise $g(x)$ could not be the monic polynomial
of lowest degree in $C$. This is where the factorization of $x^n-1$ of the last section becomes important. First let us explain what it means to generate a code by making use of a simple relationship between $g(x)$ and a special matrix well known in the theory of error correcting codes called the generator matrix. Note that we can write $g(x)=g_{0}+g_{1}x+\cdots g_{n-k}x^{n-k}$. We then can write the $k\times n$ generator matrix of the code as 
\begin{equation*}
\left( 
\begin{array}{cccccccc}
g_{0} & g_{1} & \cdots & g_{n-k} & 0 & 0 & \cdots & 0 \\ 
0 & g_{0} & \cdots & g_{n-k-1} & g_{n-k} & 0 & \cdots & 0 \\ 
0 & 0 & \cdots &  &  &  & \cdots & 0 \\ 
0 & 0 & \cdots &  & g_{0} & g_{1} & \cdots & g_{n-k}%
\end{array}
\right) .
\end{equation*} In this way, the row space of this matrix is $C$.

The previous arguments all point to the fact that if you are able to factorize $x^n-1$ into irreducible polynomials, then you can generate every cyclic code of length n over $\mathbf{F}_q$. If we can
write $x^{n}-1=w_{1}(x)w_{2}(x)\cdots w_{t}(x)$ as the decomposition of $%
x^{n}-1$ into irreducible factors, then we can generate $2^t -2$ different cyclic codes by taking any non-trivial product of the factors $w_i(x)$ as the generator polynomial. If for example you take $w_i(x)$ to be the generator polynomial, you obtain what is known as a \emph{maximal} cyclic code and if you choose $\frac{x^{n}-1%
}{w_{i}(x)}$ then you obtain an \emph{irreducible} cyclic code. It is clear that $t$ is the number of cyclotomic cosets modulo $n$. We mention one more definition pertinent to coding theory as we shall need it to understand another application of cyclotomic cosets. 

\begin{definition}
Let $C$ be a linear code of length $n$ and let $A_{i}$ be the number of
vectors in $C$ having $i$ non-zero entries (Hamming weight of $i$) . Then
the weight enumerator of $C$ is the bi-variate polynomial%
\begin{equation*}
A(x,y)=\sum_{i=0}^{n}A_{i}x^{n-i}y^{i}.
\end{equation*}%
The set $\{A_{i}\}$ is called the weight spectrum of the code.
\end{definition}

Associated with any $[n,k]$ linear code $C$ is its $[n,n-k]$ dual code $C^{\bot }$. The relation between the weight enumerator $A$ of a code 
$C$ over the field $\mathbf{F}_{q^{k}}$, and the weight enumerator $A^{\bot
} $ of the dual code $C^{\bot }$ is given by the MacWilliams identity \cite%
{Lidl:97}: 
\begin{equation}
A^{\bot }\left( x,y\right) =q^{-k^2}A\left(y-x,y+(q^k-1)x \right). \label{mac}
\end{equation}%

The computation of the weight enumerator polynomial is known to be a $\#$P-hard problem \cite{papa,Welsh}. This should not be surprising as the weight enumerator is an instance of the Tutte polynomial (as is the Jones polynomial from knot theory and the Potts partition function) \cite{Welsh}. In the next subsection we come to an overview of our last application, namely the computation of the weight enumerator polynomial.  In \cite{geraci}, cyclotomic cosets provide a speed up in the computation of the partition function for the Potts model. Given the above relationship between the weight enumerator and the partition function, it is of no surprise that cyclotomic cosets provide a \emph{practical} speed up in the evaluation of both of these functions. Again, the use of cyclotomic cosets provides no speed up asymptotically. 

\subsubsection{The Computation of Weight Enumerators for Irreducible Cyclic Codes}

We now briefly introduce characters over finite fields and Gauss sums as this will provide a vital link between quantum computation and the weights of words in a certain subset of the set of all irreducible cyclic codes. From here one will be able to see a very useful application of cyclotomic cosets. Ultimately, we wish to provide a quantum algorithm for the exact evaluation of the weight enumerator for a restricted class of codes by making use of a quantum algorithm for Gauss sums \cite{van}. This result is presented in \cite{geraci} but in the guise of evaluating the Potts partition of statistical physics. Here we make no mention of the Potts model and concentrate on the coding theoretic aspect but provide a less detailed treatment.  

Given a field $\mathbf{F}%
_{q^{k}} $, there is a multiplicative and additive group associated with it.
Namely, the multiplicative group is $\mathbf{F}_{q^{k}}^{\ast }=\mathbf{F}%
_{q^{k}}\setminus 0$ and the additive group is $\mathbf{F}_{q^{k}}$ itself.
Associated with each group are canonical homomorphisms from the group to the
complex numbers, named the additive and multiplicative characters. The
multiplicative character $\chi $ is a function of the elements of $\mathbf{F}%
_{q^{k}}^{\ast }$ and the additive character is a function of $\mathbf{F}%
_{q^{k}}$ and is parameterized by $\beta \in \mathbf{F}_{q^{k}}$.

\begin{definition}
Let $e_{\beta}$ and $\chi _{j}$ be an additive and multiplicative character
respectively. Then the Gauss Sum $G(\chi _{j},e_{\beta})$ is defined as: 
\begin{equation}
G(\chi _{j},e_{\beta})=\sum_{x\in \mathbf{F}^{\ast }}\chi
_{j}(x)e_{\beta}(x).  \label{sum}
\end{equation}
\end{definition}

 A Gauss
sum is then a function of the field $\mathbf{F}_{q^{k}}$, the multiplicative
character $\chi $ and the parameter $\beta $, and can always be written as 
\begin{equation}
G_{\mathbf{F}_{q^{k}}}(\chi ,\beta )=\sqrt{q^{k}}e^{i\gamma },  \label{gauss}
\end{equation}%
where $\gamma $ is a function of $\chi $ and $\beta $. It is in general
quite difficult to find the angle $\gamma $. The complexity of estimating
this quantity via classical computation is not known but it can be shown it is equivalent in complexity to evaluating discrete log \cite{van}.

There is a trace function over finite fields that we now define.

\begin{definition}
Let $q$ be prime, $k$ a positive integer, and let $\mathbf{F}_{q^{k}}$ be
the finite field with $q^{k}-1$ non-zero elements. The trace is a mapping $%
\mathrm{Tr}:\mathbf{F}_{q^{k}}\mapsto \mathbf{F}_{q}$ and is defined as
follows. Let $\xi \in \mathbf{F}_{q^{k}}$. Then%
\begin{equation}
\mathrm{Tr}(\xi )=\sum_{j=0}^{k-1}\xi ^{q^{j}}.  \label{eq:Tr}
\end{equation}
\end{definition}

The canonical form of an additive character is given by
\begin{equation*}
e_{\beta}(a)=e^{2\pi i/q\mathrm{Tr}(\beta a)}
\end{equation*} and the canonical form of a multiplicative character is given by
\begin{equation*}
\chi_j(\alpha^m) = e^{i\frac{2 \pi jm}{q^k-1}}
\end{equation*}
where any non-zero element in $\mathbf{F}_{q^k}$ may be written as $\alpha^m$ for some positive integer $m$, i.e., $\alpha$ is the generator of this finite field. 

We deal specifically with irreducible cyclic codes. Let $\alpha$ generate the multiplicative (cyclic) group $\mathbf{F}_{q^{k}}^{\ast }=\mathbf{F}_{q^{k}}\backslash \{0\}$.

\begin{theorem}
Each of the $q^{k}$
words of an $[n,k]$ irreducible cyclic code may be uniquely associated
with an element $\tau \in \mathbf{F}_{q^{k}}$ and may be written as
\begin{equation}
(\mathrm{Tr}(\tau),\mathrm{Tr}(\tau\alpha ^{N}),\mathrm{Tr}(\tau\alpha ^{2N}),\dots ,%
\mathrm{Tr}(\tau\alpha ^{(n-1)N})),  \label{word}
\end{equation}%
where $k$ is the smallest integer such that $q^{k}=1 \mod n$. 
\end{theorem}

For a proof of this statement see \cite{Lint:book}. 

In order to obtain $A(x,y)$ we need to find the weight spectrum $\{A_i\}$. One step in this direction is the following theorem that connects the weights of irreducible cyclic code words to Gauss sums. Let $w(x)$ be the
Hamming weight of the code word associated with $x\in \mathbf{F}%
_{q^{k}}^{\ast }$.

\begin{theorem}
\label{mceliece} \textbf{(McEliece Formula)} Let $w(\xi)$ for $\xi \in \mathbf{F}%
_{q^{k}}^{\ast }$ be the weight of the code word given by Eq.~(\ref{word}),
let $q^{k}=1+nN$ where $q$ is prime and $k$, $n$ and $N$ are positive
integers, let $d=\mathrm{gcd}(N,(q^{k}-1)/(q-1))$, and let the
multiplicative character $\bar{\chi}$ be given by $\bar{\chi}(\alpha )=\exp
(2\pi i/d)$, where $\alpha$ generates $\mathbf{F}_{q^{k}}^{\ast }$. ($\bar{%
\chi}$ is called the character of order $d$.) Then the weight of each word
in an irreducible cyclic code is given by 
\begin{equation}
w(\xi)=\frac{q^{k}(q-1)}{qN}-\frac{q-1}{qN}\sum_{a=1}^{d-1}\bar{\chi}%
(\xi)^{-a}G_{\mathbf{F}_{q^{k}}}(\bar{\chi}^{a},1).  \label{eq:Mc}
\end{equation}
\end{theorem}

For a proof of this see \cite{Berndt:book,Moisio:97}.

The main difficulty in using this theorem is that even estimating Gauss sums is computationally difficult. Fortunately, it has been shown that this is an application for which quantum computers are efficient \cite{van}. Specifically, in order to approximate $\gamma$ to within an error $\epsilon$, the computational cost is $O(\frac{1}{\epsilon }\cdot 
(\mathrm{log}(q^{k}))^2).$ 

Let us define the function 

\begin{equation}
\label{S}
S(\iota)=\frac{q^{k}(q-1)}{qN}-\frac{q-1}{qN}\sum_{a=1}^{d-1}\bar{\chi}(\alpha
^{\iota})^{-a}\sqrt{q^{k}}e^{i\widetilde{\gamma _{a}}}.
\end{equation}

This equation is just the expansion of the formula for $w(y)$ where now we take $\alpha$ to be the primitive element in $\mathbf{F}_{q^k}$ (i.e., any element in the field may be written as $\alpha^\iota$). This means that if we were able to find the range of $S(\iota)$ we would have all the weights of the corresponding code. Of course, it does look like we have to evaluate an exponential number of words in $k$, the dimension of the code. This is not the case in all situations however and this is where cyclotomic cosets will play a role. Note the following proposition.

\begin{myproposition}
\label{prop:N}In an $[n,k]$ irreducible cyclic code there are at most $N$
words of different non-zero weight where $N=(q^{k}-1)/n$.
\end{myproposition}

\begin{proof}
For any irreducible cyclic code we have the relation $q^k-1 = nN$ over the
field $\mathbf{F}_q$. The length of each word is $n$ and any cyclic
permutation of a word preserves the Hamming weight. Therefore, for each word
there are $n-1$ other words of equal weight. As there are $q^{k}-1$ words of
non-zero weight, if we assume that every word that does not arise from the
cyclic permutation of another word is of a different weight, then there are $%
(q^{k}-1)/n$ words of different weight. Being however that there is the
possibility of repeats in weight among words which are not cyclic
permutations of each other, there are at most $N$ different weights.
\end{proof}

This means that it is in fact $N$ and not $n$ which will determine the complexity of finding the weight spectrum $\{A_i\}$. The first restriction that we make on our codes is that we only consider families of codes where $N$ grows polynomially in $k$. In this way, we may claim that our algorithm for the exact evaluation of the weight enumerator is efficient as will be shown below. 

It turns out that cyclotomic cosets are a help here. This occurs because each element in a
given coset has the same value of $S(\iota)$. This is due to the fact that
the mapping $x\mapsto x^{q^j}$ is a permutation of $\mathbf{F}_{q^{k}}$ (Frobenius automorphism) and in fact this mapping is an automorphism for $\mathbf{Z}_N$ when $q$ and $N$ are relatively prime\cite{Lidl:97}.  Let us assume that we have all $d-1$ Gauss sums necessary to compute $S(\iota)$ (via a quantum computation for example). Let us call these Gauss sums $\Lambda_a$.  We then must convince ourselves that $S(g) = S(f)$ whenever $g=fq^j$ for some integer $j$. We have 

\[ S(g) \sim \sum_{a=1}^{d-1} \left ( e^{\frac{2 \pi i fq^j}{d}} \right )^{-a} \Lambda_a = \sum_{a=1}^{d-1} \left ( \left ( e^{\frac{2 \pi i f}{d}} \right )^{-a} \right )^{q^j}\Lambda_a . \] One can show that $\mathrm{gcd}(q^j,d)=1$ and therefore  
the mapping \[\left ( e^{\frac{2\pi i f}{d}} \right) \mapsto \left ( e^{\frac{2\pi i f}{d}} \right)^{q^j} \] is just a permutation of the cyclic group of order $d$ generated by the primitive root of unity, i.e., the above mapping is an automorphism. This means that the sum does not change and therefore we have that $S(g)=S(f)$. This means that $S(\iota)$ is invariant over individual cosets.

It is known that the number of cyclotomic cosets is equal to 
\begin{equation}
N_{C}=\sum_{f|N}\frac{\phi (f)}{\mathrm{ord}_{q}f}  \label{N_C}
\end{equation}
where $\phi (f)$ is the Euler totient (the number of positive integers which
are relatively prime to $f$ and $s=\mathrm{ord}_{q}f$ means that $s$ is the
smallest positive integer such that $q^{s}=1\mod f.$)\cite{Lidl:97} There are many instances where $N_C << N$ but asymptotically, it does not make an exponential difference. However the difference can be significant. Take for example $N=358701$. The number of 2-cyclotomic cosets is 546. One can clearly see that this has the potential for a large speed up for the task of evaluating weight enumerators.

\section{A Theorem on the Exact Evaluation of the Weight Enumerator via Quantum Computation for a certain class of Cyclic Codes}

Earlier, we made mention of a theorem presented in \cite{van} that gives a poly-logarithmic algorithm for the estimation of a Gauss sum. The algorithm is for an approximation of the angle $\gamma$ in 
\begin{equation}
G_{\mathbf{F}_{q^{k}}}(\chi ,\beta )=\sqrt{q^{k}}e^{i\gamma },  \label{gauss}
\end{equation}%
up to an error $\epsilon$. This means that if $\gamma_a$ is the actual angle then the quantum algorithm returns $\gamma$ such that $|\gamma_a - \gamma|< \epsilon$. The smaller we wish to make $\epsilon$ the more times we would have to run our quantum algorithm, i.e., if we want $\epsilon$ accuracy we have to run the algorithm $1/\epsilon$ times. How can we use this result to obtain the exact weight spectrum? Clearly the error would propagate when we attempted to find the range of $S(\iota)$. This is dealt with in the paper \cite{geraci} but we give a brief review. 

Fortunately, there is a theorem which gives us some information about the weights of words in irreducible cyclic codes. 

\begin{theorem}
\label{mceliece2} \textbf{{(McEliece \cite{aubry})}} All the weights of an $%
[n,k]$ irreducible cyclic code are divisible by $q^{\theta_{n,k} -1}$, where 
$\theta_{n,k}$ is given by \begin{equation}
\theta _{n,k}=\frac{1}{q-1}\min_{0<j\leq \alpha k^{s(k)}}S^{\prime }(jn)
\label{eq:theta}
\end{equation}%
(where $S^{\prime }(x)$ is the sum of the digits of $x$ in base $q$)
\end{theorem}

Being that the weights are integers, this theorem gives us a clue as to the distance between weights. What this means is that if we can make $\epsilon$ small enough, we will be able to guarantee that the range of $S(\iota)$ are the actual weights even though we are using an approximation of the Gauss sum. In \cite{geraci} it was shown that \[ \epsilon \leq \frac{q^{\theta _{n,k}-1}}{4\sqrt{q^{k}}} \] is sufficient. Further, it can be shown that for any fixed $\epsilon<1$, there is a family of cyclic codes which conform to the necessary restrictions required to obtain the exact weight enumerator. There is a polynomial speed up in the dimension $k$ and an exponential speed up in $q$ over the best classical algorithms. See \cite{geraci,vlugt,Moisio:99} for details. For completeness we mention justification for this claim of algorithmic speed up. Note that in \cite{Moisio:99}, they give an algorithm for computing the weight distribution of binary \emph{index 2} irreducible cyclic codes. The algorithm is efficient and is due to the fact that there is an efficient way of solving the Diophantine equation necessary for this case. As indicated in \cite{vlugt}, the weight distributions of irreducible cyclic codes are intimately related to Gauss sums (as these functions are related to the number of rational points on Hasse-Davenport curves). Thus, for the index 2 cases explored in \cite{Moisio:99}, they used a special form that Gauss sums take for this situation as well as information from the solution of the particular Diophantine equation. Now, \emph{index 2} refers to the fact that the dimension $k$ of the code is equal to $\phi(N)/2$, where $\phi$ is the Euler totient  function. Asymptotically it is well known that $N^{1-\epsilon} < \phi(N) < N$, and thus we essentially have $k \sim N$. This means that the situations that we are able to handle are computationally much more difficult to deal with than these situations and the quantum computers ability to approximate Gauss sums provides a very significant advantage. In fact, the assumption that the length of of the codes considered in this paper grow exponentially with $k$, makes it very unlikely that any approach devoid of computations of Zeta functions or Gauss sums will be sufficient.

We now give a formal definition for the class of codes for which this applies and a theorem that summarizes the results. 

\begin{definition}
 Given a constant $\epsilon <1$, $\mathrm{ICQ}_{\epsilon }$ is the class of irreducible cyclic codes of dimension $k$ and length $n$, such that 
\begin{equation}
  n=\frac{q^{k}-1}{\alpha k^{s}}
  \label{eq:n}
\end{equation}%
(where $\alpha \in \mathbf{R}$ is chosen so that $n\in \mathbf{N}$ and where 
$s \in \mathbf{R}$ determines the complexity and the instances of codes considered) and 
\begin{equation}
\theta _{n,k}=\frac{1}{q-1}\min_{0<j\leq \alpha k^{s}}S^{\prime }(jn)
\label{eq:theta}
\end{equation}%
(where $S^{\prime }(x)$ is the sum of the digits of $x$ in base $q$) so that 
\begin{equation}
\epsilon \leq \frac{q^{\theta _{n,k}-1}}{4\sqrt{q^{k}}}.
\end{equation}
$ICQ_\epsilon$ also includes the cyclic $[n,n-k]$ dual codes and all equivalent codes \cite{Lint:book}.
\end{definition}

\begin{theorem}
\label{th 1} A quantum computer can return the exact weight enumerator polynomial $A(x,y)$ for codes in $\mathrm{ICQ}_{\epsilon }$. For each family $\mathrm{ICQ}_{\epsilon }$ ($\epsilon$ fixed), the overall running time is $
O(k^{2s}(\log q)^{2})$ and the success probability is at least $%
1-\delta $, where $\delta =[2((q^{k}-1)^{2}\epsilon -2)]^{-1}$.
\end{theorem}

This theorem imposes a restriction on the fundamental relationship $nN=q^k-1$ in that we impose that asymptotically $N=O(k^s)$. This essentially means that we consider codes for which the lengths of the codes grow exponentially. This is a good restriction for it makes brute force classical computation not feasible. We do not supply a proof for the theorem as it is essentially the same as the proof given in \cite{geraci}. We do however supply an overview of the algorithm for computing the weight enumerator of a code in $\mathrm{ICQ}_{\epsilon }$. The success probability comes from the fact that the evaluation does depend on a quantum algorithm and thus is ultimately probabilistic. See \cite{geraci,van, Nielsen:book} for details.

\subsection{Overview of the Algorithm to Obtain the Exact Weight Enumerator of a Code in $\mathrm{ICQ}_{\epsilon }$}

In \cite{geraci}, a quantum algorithm for checking whether graphs are members of the family $\mathrm{ICCC}_\epsilon$ is given where it essentially checks whether codes are members of $\mathrm{ICQ}_{\epsilon }$ as defined in this paper. This quantum algorithm is exponentially faster than the best classical algorithm for it requires the computation of the discrete logarithm \cite{Shor}. We do not present it here but instead just assume that we know that a code does indeed belong to $\mathrm{ICQ}_{\epsilon }$

\begin{enumerate}

\item Let $N=O(k^{s})$ where $s$ is a constant integer that determines the
complexity of the algorithm. Take $C$ as our irreducible cyclic code
of length $n=\frac{q^{k}-1}{N}$ and dimension $k$ (or the dual code).

\item Find the $q$-cyclotomic cosets of $\{0,1,\dots ,N-1\}$. This step
requires at most linear time in $N$. (See the next section)

\item Using the quantum algorithm for Gauss sums \cite{van} we are be able
to estimate the weights of the words.  Use the Gauss sum
algorithm to return the phases $\gamma _{1},\dots ,\gamma _{d-1}$ [Eq. (\ref%
{gauss})] and then input these values into the function $S(\iota)$. According to the McEliece Formula (Th.~\ref{mceliece}) we have to make $%
d-1$ (where $d=\gcd (N,\frac{q^{k}-1}{q-1})$) calls to the quantum oracle
and we can use these evaluations for each representative $i$ of the $q$%
-cyclotomic cosets of $\{0,1,\dots ,N-1\}$. This step has time complexity $%
O(dk^{2}(\log q)^{2})$ \cite{van,geraci}.

\item Let $b_{1},b_{2},\dots ,b_{N_{C}}$ be the coset representatives from the $N_{C}$ cosets. Now each coset has cardinality $%
v_{i}$, i.e., $b_{i}$ belongs to coset $i$ which has $v_{i}$ elements. We
evaluate $\omega _{i}=S(b_{i})$ for each $b_{i}$, remembering that each $%
\omega _{i}$ occurs $v_{i}$ times. We end up with a list $(\omega
_{1},\omega _{2},\dots ,\omega _{N_{C}})$ as well as a list $%
(v_{1},v_{2},\dots ,v_{N_{C}})$ of multiplicities. This step will have an $O((d-1)\cdot N_C)$ time cost.

\item Now perform a tally of repeats of the $\omega _{i}$ for each $i\in
\{1,...,N_{C}\}$. This returns a set of indices $\Lambda _{i}\equiv
\{j_{i}\}\subseteq \{1,...,N_{C}\}$. We add the corresponding $v_{j_{i}}$
which yields $a_{i}=\sum_{j\in \Lambda _{i}}v_{j}$, the number of words of
weight $\omega _{i}$ up to cyclic permutations. To account for cyclic
permutations due to the fact that we are working over cyclic codes, we have $%
A_{i}=na_{i}$, which is the desired weight spectrum. The tally will have an $O(\sqrt{N_C})$ time cost using Grover's quantum search algorithm \cite{Grover:96}. (This will have no affect on the overall complexity.)

\item Combining the previous steps, we now have determined the weight spectrum $A_{i}$ in time $
O(k^{2s}(\log q)^{2})$ (by modestly taking $N_C = O(k^s)$, i.e., essentially ignoring the contribution of the cyclotomic cosets).  This means that we have the coefficients for $A(x,y)$ as well as the exponents and thus, are done.

\end{enumerate}

\section{A Classical Algorithm for the Computation of Coset Leaders and Coset Size}

The algorithm for the calculation of the cyclotomic cosets themselves is quite simple; it is essentially a sieve method of the kind commonly used in number theoretic algorithms such as those for prime factorization.

\begin{tabbing}
\textsc{CosetLeaders} ($N$, $p$) \\
xxxx \= xxxx \= xxxx \= xxxx \= xxxx \= \kill
\> \texttt{Array} $A$ (\texttt{size} $N$), \texttt{initialize to} $unmarked$ \\
\> \texttt{for} $i = 0$ \texttt{to} $N-1$ \texttt{do} \\
\> \> \texttt{if} $A_i = unmarked$ \texttt{do} \\
\> \> \> \texttt{output} ``New coset leader $= i$'' \\
\> \> \> $a \leftarrow i$, $s \leftarrow 0$ \\
\> \> \> \texttt{while} $A_a = unmarked$ \texttt{do} \\
\> \> \> \> \texttt{mark} $A_a$ \\
\> \> \> \> \texttt{increment} $s$ \\
\> \> \> \> $a \leftarrow a \times p \ (\textrm{mod } N)$ \\
\> \> \> \texttt{end while} \\
\> \> \> \texttt{output} ``Coset size $= s$'' \\
\> \> \texttt{end if} \\
\> \texttt{end for} \\
\texttt{end} \textsc{CosetLeaders} \\
\end{tabbing}

The outer loop scans for coset leaders, which here are unmarked numbers of the form $ap^0$, while the inner loop sieves out other coset members i.e. $ap^k$ for $k=1$ to $s-1$, where $s$ is the size of a particular coset. Since, as explained in section II, the cosets partition $1$ to $N-1$, and $s$ is the smallest integer such that $a(p^s - 1) \equiv 0 \ (\textrm{mod } N)$, on termination the inner loop has returned to the original coset leader $ap^0$ after marking every other member.

While the algorithm features nested loops, its running time is linear in $N$, since the inner loop is activated only once per coset, and the number of iterations for a particular coset are equal to the size of that coset. In fact, it is easy to see that every element in $A$ is read only twice (once in an unmarked state, and once in a marked state) and of course marked only once (as well as unmarked once, during initialization). It should be noted that while the algorithm is soft-$\mathcal{O}(N)$, in terms of general complexity it is not polynomial with respect to the input size, but only \em pseudo-polynomial\em, since $N$ and $p$ are given as (presumably) binary numbers. This is of course the best that can be done for enumeration problems of this sort, which have very succinct inputs consisting of only 1 or 2 numbers but outputs that consist of relatively long lists (the number of cosets can approach $\frac{N}{2}$, as in the example given in section II). As well, like other sieve algorithms, the storage requirements can be a bit onerous for large $N$, but this can be helped a bit by doing things such as implementing $A$ as a bit-array. Such optimizations make the problem feasible for $N$ up to several billion on one of today's ordinary household computers.

\begin{acknowledgments}
This work was done under the support of ARO grant W911NF-05-1-0440 (to D.A. Lidar). I 
would also like to thank D.A. Lidar and Marko Moisio for helpful conversations. 
\end{acknowledgments}

\section{Literature Citations}

\end{document}